\documentstyle[12pt,epsfig]{article}
\newcommand\slv{v\kern-5pt\raise1pt\hbox{$\scriptstyle/$}\kern1pt}

\newcommand{\be}{\begin{equation}}
\newcommand{\ee}{\end{equation}}
\begin{document}
\thispagestyle{empty}
\begin{flushright}
WUE-ITP-97-051\\
\end{flushright}
\vspace{0.5cm}
\begin{center}
{\Large \bf Form Factors of $\gamma^{*}\rho\to\pi$ and 
 $\gamma^{*}\gamma\to\pi^0$}\\
\vspace{0.2 cm} 
{\Large \bf Transitions  and Light-Cone Sum Rules}
\vspace{1.7cm}

{\bf Alexander ~Khodjamirian $^*$\\[1cm]
\em Institut f\"ur Theoretische Physik, Universit\"at W\"urzburg, D-97074 W\"urzburg, Germany} \\
\end{center}
\vspace{2cm}
\begin{abstract}\noindent
{ The method of light-cone QCD sum rules is applied to 
the calculation of  the form factors of 
$\gamma^*\rho \to \pi$ and $\gamma^*\gamma \to \pi^0$ transitions. 
We consider the dispersion relation for 
the $\gamma^*(Q^2)\gamma^*(q^2) \to \pi^0$ amplitude in
the variable $q^2$. At large virtualities $q^2$ and $Q^2$,
this amplitude is calculated in terms of light-cone wave functions 
of the pion. As a next step, the light-cone 
sum rule for the $\gamma^*(Q^2) \rho \to \pi $ form factor is derived. 
This sum rule, together with  the quark-hadron duality, provides 
an estimate of the hadronic spectral density 
in the dispersion relation. Finally, the 
$\gamma^*(Q^2)\gamma \to \pi^0$ form factor is obtained 
taking the $q^2=0$ limit in this relation. 
Our predictions are valid at $Q^2 \geq 1 ~\mbox{GeV}^2$
and have a correct asymptotic behaviour at large $Q^2$.}    
\end{abstract}

\bigskip
\begin{center}
To appear in Eur. Phys. J. C
\end{center} 
\vspace*{\fill}
\noindent $^*${\small \it On leave from 
Yerevan Physics Institute, 375036 Yerevan, Armenia } \\
\newpage

\section {Introduction}
Light-cone wave functions (distribution amplitudes) of hadrons 
have been introduced in QCD to define the long-distance part of 
exclusive processes with large momentum transfer \cite{BL,ER}. 
The same  wave functions serve as an input in QCD light-cone 
sum rules \cite{BBK,BF,CZ90,BKR,BBKR,BH} which are based on the  
light-cone operator product expansion (OPE) of vacuum-hadron correlators.  
At asymptotically large normalization scale, the light-cone 
wave functions are given by perturbative QCD. To estimate or at least 
to constrain nonasymptotic corrections, one needs either 
nonperturbative methods or, in a more direct way, 
measurements of hadronic quantities  which 
are sensitive to the shape of light-cone wave functions.

One of the simplest processes determined by the light-cone wave functions
of the pion is  the transition  
$
\gamma^{*}(q_1)\gamma^{*}(q_2)\to \pi^0(p)~
$
of two virtual photons into a neutral pion. This process is 
defined by  the matrix element 
\be
\int d^4x e^{-iq_1x}\langle \pi^0 (p)\mid
T\{ j_\mu(x) j_\nu(0)\}\mid 0\rangle = i\epsilon_{\mu\nu\alpha\beta}
q_1^\alpha q_2^\beta F^{\gamma^{*}\pi}(Q^2,q^2)~,
\label{ampl}
\ee
where $Q^2=-q_1^2$, $q^2=-q_2^2~$ are the virtualities 
of the photons and $j_\mu=(\frac23\bar{u}\gamma_\mu u -
\frac13\bar{d}\gamma_\mu d)$ is the quark electromagnetic current.
If both $Q^2$ and $q^2$ are sufficiently large,  
the $T$-product of currents in (\ref{ampl}) can be expanded 
near the light-cone $x^2=0$. The leading term of this expansion
yields \cite{BL}: 
\be
F^{\gamma^{*}\pi}(Q^2,q^2)= \frac{\sqrt{2}f_\pi}{3}\int\limits_0^1 
\frac{du~\varphi_\pi(u)}{Q^2(1-u) +q^2u}~,
\label{leading}
\ee
where $\varphi_\pi(u)$ is the pion wave function of twist 2. 
Nonleading terms of the light-cone OPE 
are determined by pion wave functions of higher twist.
Their contributions to $F^{\gamma^{*}\pi}$ 
are suppressed by additional inverse powers of  photon virtualities.
Therefore, measurements of the  
form factor $F^{\gamma^{*} \pi}(Q^2,q^2)$ 
at large $Q^2$ and $q^2 \neq Q^2$ will be a direct source of 
information on  $\varphi_{\pi}(u)$. 

Recently, the CLEO collaboration has measured \cite{CLEO}
the photon-pion transition form factor 
$F^{\gamma\pi}(Q^2)\equiv F^{\gamma^{*}\pi}(Q^2,0)$, where 
one of the photons is nearly on-shell and the other one is 
highly off-shell, with the virtuality 
in the range $1~ \mbox{GeV}^2 < Q^2 < 10~ \mbox{GeV}^2$.
A straightforward calculation of $F^{\gamma\pi}(Q^2)$
in QCD is, however, not possible. In particular, 
at $q^2\to 0$, it is not sufficient to retain 
a few terms of the light-cone OPE of (\ref{ampl}). 
One has, in addition, to take into account the interaction 
of the small-virtuality photon at long distances 
of $O( 1/\sqrt{q^2})$ (for a recent discussion, see \cite{RR,RM}). 

In this paper, a simple method is suggested to calculate the form factor 
$F^{\gamma \pi}(Q^2)$ at sufficiently large $Q^2$
(practically, at $Q^2 \geq 1$ GeV$^2$), in terms 
of the pion light-cone wave functions.  
The  method allows to avoid the problem of 
the photon long-distance interaction by performing  
all QCD calculations at sufficiently large $q^2$. 
In parallel, the form factor of the $\gamma^* \rho \to \pi$ transition 
is obtained from the light-cone sum rule. 
In the following sections, the calculational procedure is described, 
the light-cone OPE of the amplitude (\ref{ampl})
is performed up to twist 4 and the numerical results for the 
$\gamma^*\rho \to \pi$ and $\gamma^*\gamma \to \pi^0$ transition
form factors are presented and discussed. The account of 
$ O(\alpha_s)$ corrections is postponed to a future study.   

\section{The method} 
Our starting object is the dispersion relation
for the amplitude $F^{\gamma^{*}\pi}(Q^2,q^2)$ in the variable $q^2$
and at fixed large $Q^2$. Physical states in the $q^2$--channel include  
vector mesons $\rho,\omega,\rho',\omega',...$ and a continuum
of hadronic states with the same quantum numbers. 
We assume that the spectral density
in the dispersion relation can be approximated 
by the ground states $\rho$, $\omega$ and the higher
states with an effective threshold $s_0$:
\be
F^{\gamma^{*}\pi}(Q^2,q^2) 
=\frac{\sqrt{2}
f_\rho F^{\rho \pi}(Q^2)}{m_{\rho}^2+q^2}
+ \int\limits_{s_0}^\infty ds~ \frac{\rho^h(Q^2,s)}{s+q^2}.  
\label{disp}
\ee
Here, the $\rho$ and $\omega$  contributions
are combined in one resonance term 
assuming $m_\rho \simeq m_\omega$, adopting zero-width approximation and 
defining the matrix elements  of electromagnetic transitions 
\be
\frac13\langle \pi^0(p)\mid j_\mu\mid \omega(q_2)\rangle \simeq
\langle \pi^0(p)\mid j_\mu\mid \rho^0(q_2)\rangle =
F^{\rho \pi}(Q^2)
m_\rho^{-1}\epsilon_{\mu\nu\alpha\beta}e^{\nu}q_1^\alpha
q_2^\beta 
~, 
\label{formf}
\ee
and the decay constants of vector mesons
\be
3\langle \omega \mid j_\nu \mid 0 \rangle \simeq
\langle \rho^0 \mid j_\nu \mid 0 \rangle = 
\frac{f_\rho}{\sqrt{2}} m_\rho e^{ *}_\nu~, 
\label{formrho}
\ee
$e_\nu$ being the polarization vector of the $\rho$-meson. 
Approximate relations in (\ref{formf}) and (\ref{formrho})
follow from the quark content of $\omega$ and $\rho$ 
and from the isospin symmetry. 

Two important points should be emphasized. First, the dispersion relation 
(\ref{disp}) does not contain  subtraction terms \cite{ShifSmil}. 
Otherwise, at $q^2\to \infty$, the asymptotic
behaviour of $F^{\gamma^{*} \pi}(Q^2,q^2)$ 
dictated by (\ref{leading}) will be violated. 
Second, due to absence of massless resonances, 
it is possible to analytically continue (\ref{disp}) to 
$q^2=0$. One then obtains the form factor 
$F^{\gamma \pi}(Q^2)$. 
Therefore, the outlined problem 
can be solved if  the function $F^{\rho\pi}(Q^2)$ and the integral 
over $\rho^h(Q^2,s)$ in the dispersion relation (\ref{disp}) are known
\footnote{ A similar approach was used 
in \cite{GIKO} to estimate the structure function of the real photon.
The dispersion relation for the structure function 
of the virtual photon was analytically continued 
to the zero virtuality limit.}. 

To estimate the spectral density $\rho^h(Q^2,s)$ of the higher states 
in (\ref{disp}), we employ the usual quark-hadron duality: 
\be
\rho^h(Q^2,s) = \frac{1}{\pi}\mbox{Im} F^{\gamma^*\pi}_{QCD}(Q^2,s)
\Theta(s-s_0)\label{duality}~,
\ee 
where $F^{\gamma^*\pi}_{QCD}$ is the amplitude (\ref{ampl})
calculated in QCD using the light-cone OPE.  
The form factor of the 
$\gamma^*\rho \to \pi$ transition determining the residue of 
the resonance term in (\ref{disp}) can also be obtained
in the same framework.
We follow the procedure described in \cite{BKR,BBKR} and  \cite{BH} 
where the $B\to\pi$ form factor and the pion electromagnetic
form factor have been calculated, respectively.  
One equates the dispersion relation 
(\ref{disp}) with $F^{\gamma^{*}\pi}_{QCD}(Q^2,q^2)$
at large  $q^2$, where the light-cone OPE   
is reliable and  higher-twist terms are under quantitative control: 
\be
\frac{\sqrt{2}f_\rho F^{\rho \pi}(Q^2)}{m_{\rho}^2+q^2}
+ \int\limits_{s_0}^\infty ds~ \frac{\rho^h(Q^2,s)}{s+q^2}  
= \frac{1}{\pi}\int\limits_0^\infty ds~\frac{\mbox{Im}F^{\gamma^{*}\pi}_{QCD}
(Q^2,s)}{s+q^2}~.  
\label{dispQCD}
\ee
Using (\ref{duality}), subtracting the integral over $\rho^h(Q^2,s)$ 
from the dispersion integral on the r.h.s. of (\ref{dispQCD})
and performing the Borel transformation in $q^2$ yields the 
light-cone sum rule 
\be
\sqrt{2}f_\rho~F^{\rho \pi}(Q^2)=
\frac1{\pi}\int\limits_{0}^{s_0} ds ~
\mbox{Im}F^{\gamma^{*}\pi}_{QCD}(Q^2,s)\exp\left( \frac{m_\rho^2-s}{M^2}\right)~.
\label{lcsr}
\ee
Substituting (\ref{lcsr}) and the duality  
approximation (\ref{duality}) in the initial dispersion relation (\ref{disp}), 
and, finally, taking the $q^2 \to 0$ limit we obtain 
an estimate of the $\gamma^*\gamma\to \pi^0$ 
form factor:  
\be
F^{\gamma\pi}(Q^2) = 
\frac{1}{\pi m_\rho^2}\int\limits_{0}^{s_0} ds~
\mbox{Im}F^{\gamma^{*}\pi}_{QCD}(Q^2,s)
\exp\left( \frac{m_\rho^2-s}{M^2}\right)+\frac1{\pi}
\int\limits_{s_0}^\infty 
\frac{ds~}{s}\mbox{Im}F^{\gamma^{*}\pi}_{QCD}(Q^2,s) ~.
\label{answer}
\ee

\section{ Light-cone expansion}
It remains to calculate 
the amplitude $F^{\gamma^*\pi}(Q^2,q^2)$ using light-cone OPE 
and to take its imaginary part. The procedure essentially 
follows \cite{BF,BKR,BBKR,BH} where vacuum-pion
correlators similar to the amplitude (\ref{ampl}) have been calculated  
\footnote{The light-cone OPE of the amplitude
(\ref{ampl}) was also studied in \cite{Gorsky} using 
different definitions of the higher-twist  wave functions.}.

To obtain the contribution of two-particle (quark-antiquark)
wave functions of the pion, one has to contract 
two quark fields in the product of currents in (\ref{ampl}): 
$$
\int d^4x e^{-iq_1x}\langle \pi^0 (p)\mid
T\{ j_\mu(x) j_\nu(0)\}\mid 0\rangle 
= 2\int d^4x e^{-iq_1x} 
$$
\be
\times\langle \pi^0 (p)\mid\ \left( \frac23\right)^2 \bar{u}(x)\gamma_\mu i\hat{S}_u(x,0)
\gamma_\nu u(0) + \left( \frac13\right)^2\bar{d}(x)\gamma_\mu i\hat{S}_d(x,0)
\gamma_\nu d(0)
\mid 0\rangle ~,
\label{fig1}
\ee
and substitute the free-quark propagator 
\be
i\hat{S}^0_q(x,0)
=\langle 0 \mid T\{q(x)\bar{q}(0)\} \mid 0\rangle = 
\frac{i\not\!x}{2\pi^2x^4}.
\label{prop}
\ee
The light quark masses and the pion mass  are neglected in this 
calculation ($p^2=m_\pi^2 \simeq 0$). 
The approximation (\ref{fig1}) 
corresponds to the diagram shown in Fig. 1a.
The factor 2 takes into account 
two equal contributions of this 
diagram with opposite directions of quark lines.
The matrix elements of nonlocal quark-antiquark operators 
emerging in (\ref{fig1}) are expanded near the light-cone: 
$$
\langle\pi^0(p)|\bar{u}(x)\gamma_\mu\gamma_5u(0)|0\rangle =
-\langle\pi^0(p)|\bar{d}(x)\gamma_\mu\gamma_5d(0)|0\rangle 
$$
\be
=-ip_\mu \frac{f_\pi}{\sqrt{2}}\int_0^1du\,e^{iup\cdot x}
\left(\varphi_\pi (u)+x^2g_1(u)\right)
+
\frac{f_\pi}{\sqrt{2}}\left( x_\mu -\frac{x^2p_\mu}{p \cdot x}\right)\int_0^1
du\,e^{iup \cdot x}g_2(u)~,
\label{phi}
\ee
where $\varphi_\pi(u)$ and $g_{1,2}(u)$ are the twist 2 
and twist 4  wave functions of the pion, respectively. 
The twist 3 terms of the light-cone OPE of the amplitude
(\ref{ampl}) are proportional to $m_\pi^2$ and therefore vanish in the adopted
chiral limit. Terms corresponding  to twists higher than 4 are neglected. 
The light-cone gauge is assumed for the gluon
field suppressing the path-ordered 
gauge factors in the matrix elements (\ref{phi}).
To twist 4 accuracy, the result for the diagram Fig. 1a reads:
\be
F^{\gamma^{*} \pi}_{(a)}(Q^2,q^2)=
\frac{\sqrt{2}f_\pi}3 \left(\int\limits_0^1 
\frac{du~\varphi_\pi(u)}{Q^2(1-u) +q^2u}-
4\int\limits_0^1 \frac{du ~(g_1(u)+ G_2(u))}{(Q^2(1-u) +q^2u)^2}\right),
\label{opea}
\ee
where $G_2(u) = -\int_0^u dv~ g_2(v)$.
The first, leading term was already given in (\ref{leading}).

Furthermore, there are contributions to the light-cone OPE 
due to many-particle (higher Fock) states in the pion. 
With the same accuracy, one has to include the
quark-antiquark-gluon wave functions
taking into account the gluon emission from the virtual 
quark (Fig. 1b). In order to obtain this contribution,
the quark propagator including the interaction 
with gluons in first order: 
\be
i\hat{S}^G_q(x,0)= -\frac{ig_s}{16\pi^2x^2}\int\limits ^1_0 dv 
\left( \not\!x\sigma_{\alpha\beta} -4ivx_\alpha\gamma_\beta
\right)G^{\alpha\beta}(vx)~
\label{firstG}
\ee
with $G_{\alpha\beta} = G^a_{\alpha\beta}\frac{\lambda^a}2$,
should be substituted in (\ref{fig1}). One then encounters 
matrix elements 
of nonlocal quark-antiquark-gluon operators.
They are defined in \cite{BF,BF2}:
\begin{eqnarray}
\lefteqn{
\langle\pi^0(p) |\bar{u}(x)\gamma_\mu\gamma_5 g_sG_{\alpha\beta}(vx)u(0)|0\rangle =
-\langle\pi^0(p) |\bar{d}(x)\gamma_\mu\gamma_5 g_sG_{\alpha\beta}(vx)d(0)|0\rangle =}
\nonumber
\\
&=&\frac{f_\pi}{\sqrt{2}}\Bigg\{\left[ p_\beta\left( g_{\alpha\mu}-\frac{x_\alpha p_\mu}{p\cdot x}\right) -
p_\alpha\left(g_{\beta\mu}-\frac{x_\beta p_\mu}{p\cdot x}\right)\right]
\int{\cal D}\alpha_i\varphi_\perp (\alpha_i)e^{ip\cdot x(\alpha_1+v\alpha_3)}\hspace{1.5cm}{}
\nonumber
\\
&& {}+\frac{p_\mu}{p\cdot x}(p_\alpha x_\beta -p_\beta x_\alpha )
\int{\cal D}\alpha_i\,\varphi_\parallel (\alpha_i)e^{ip \cdot x(\alpha_1+v\alpha_3)}\Bigg\} \,,
\label{30}
\end{eqnarray}
\begin{eqnarray}
\lefteqn{
\langle\pi^0(p) |\bar{u}(x)\gamma_\mu g_s\tilde{G}_{\alpha\beta}(vx)u(0)|0\rangle= -
\langle\pi^0(p) |\bar{d}(x)\gamma_\mu g_s\tilde{G}_{\alpha\beta}(vx)d(0)|0\rangle=}
\nonumber
\\
&=&\frac{if_\pi}{\sqrt{2}}\Bigg\{\left[ p_\beta\left( g_{\alpha\mu}-\frac{x_\alpha p_\mu}{p \cdot x}\right) -
p_\alpha\left( g_{\beta\mu}-\frac{x_\beta p_\mu}{p \cdot x}\right)\right]
\int{\cal D}\alpha_i\,\tilde{\varphi}_\perp (\alpha_i)e^{ip \cdot x(\alpha_1+v\alpha_3)}
\hspace{1.4cm}{}
\nonumber
\\
&&{}+\frac{p_\mu}{p \cdot x}(p_\alpha x_\beta -p_\beta x_\alpha )
\int{\cal D}\alpha_i\,\tilde{\varphi}_\parallel (\alpha_i)
e^{ip \cdot x(\alpha_1+v\alpha_3)}\Bigg\}\,,
\label{31}
\end{eqnarray}
where $\tilde{G}_{\alpha\beta}= \frac12
\epsilon _{\alpha\beta \sigma\tau}G^{\sigma\tau}$  and 
${\cal D}\alpha_i= d\alpha_1 d\alpha_2 d\alpha_3 \delta(1-\alpha_1-\alpha_2
-\alpha_3)$. The wave functions 
$\varphi_\perp$, $\varphi_\parallel$,
$\tilde{\varphi}_\perp$ and $\tilde{\varphi}_\parallel$ have twist 4. 
Using (\ref{30}) and (\ref{31})
and integrating (\ref{fig1}) over $x$, 
one obtains the answer for the diagram Fig. 1b:
\begin{eqnarray}
F^{\gamma^{*} \pi}_{(b)}(Q^2,q^2)
=-\frac{\sqrt{2}f_\pi}3
\int\limits_0^1 \frac{du}{(Q^2(1-u) +q^2u)^2}
\int\limits^u_0~d\alpha_1~\int\limits^{1-u}_0\frac{d\alpha_2}{
\alpha_3}
\nonumber
\\
\times\left( 
\frac{1-2u+\alpha_1-\alpha_2}{\alpha_3}
\varphi_{\parallel}(\alpha_1,\alpha_2,\alpha_3)
-\widetilde{\varphi}_{\parallel}(\alpha_1,\alpha_2,\alpha_3)
\right )_{\alpha_3=1-\alpha_1-\alpha_2}~.
\label{opeb}
\end{eqnarray}
The wave functions $\varphi_\perp$ and 
$\tilde{\varphi}_\perp$ drop out due to the antisymmetry 
of the amplitude (\ref{ampl}) in $\mu,\nu$. 

Taking the sum 
of (\ref{opea}) and (\ref{opeb}) 
and transforming the integration variable,  $u \to Q^2/(s+Q^2)$, 
one obtains 
the OPE result for the amplitude (\ref{ampl}) in the
form of a dispersion integral:
$$
F^{\gamma^{*} \pi}_{QCD}(Q^2,q^2)=
\frac{\sqrt{2}f_\pi}3 \int\limits_0^1 
\frac{du}{Q^2(1-u) +q^2u}\left(\varphi_\pi(u) -
\frac{\varphi^{(4)}(u)}{Q^2(1-u) +q^2u}\right)
$$
\be
= \frac{1}{\pi}\int\limits_0^\infty ds~
\frac{\mbox{Im}F^{\gamma^{*}\pi}_{QCD}
(Q^2,s)}{s+q^2}
\label{Ffinal}
\ee
with the imaginary part
\be
\frac1{\pi}\mbox{Im}F^{\gamma^{*} \pi}_{QCD}(Q^2,s)=
\frac{\sqrt{2}f_\pi}{3}\left(\frac{\varphi_\pi(u)}{s+Q^2}
-\frac1{Q^2}\frac{d\varphi^{(4)}(u)}{ds}\right)_{u=\frac{Q^2}{s+Q^2}} ~,
\label{rhoqcd}
\ee
where the following combination of twist 4 wave functions
is introduced:
$$
\varphi^{(4)}(u)=  4\left(g_1(u)+G_2(u)\right) 
$$
\be
+ 
\int\limits^u_0~d\alpha_1~\int\limits^{1-u}_0\frac{d\alpha_2}{\alpha_3}
\left( 
\frac{1-2u+\alpha_1-\alpha_2}{\alpha_3}
\varphi_{\parallel}(\alpha_1,\alpha_2,\alpha_3)
-\widetilde{\varphi}_{\parallel}(\alpha_1,\alpha_2,\alpha_3)\right)_{\alpha_3=
1-\alpha_1-\alpha_2}~.
\label{tw4wf}
\ee

The twist 2 wave function can be expanded \cite{BL,ER}
in Gegenbauer polynomials $C^{3/2}_n$: 
 \begin{equation}
\varphi_\pi(u,\mu) = 6 u(1-u)\Big[1+ \sum_{n=2,4,...}a_n(\mu)C^{3/2}_n(2u-1)
\Big]\,,
\label{phipi}
\end{equation}
Nonperturbative effects are contained in the 
coefficients $a_n$ which logarithmically depend on the 
normalization scale $\mu$ of the wave function. 
Substituting in (\ref{tw4wf}) 
the asymptotic twist 4 wave functions from \cite{BF2}
we obtain a simple expression: 
\be
\varphi^{(4)}(u,\mu)= \frac{80}3 \delta^2(\mu) u^2(1-u)^2~,
\label{phi4}
\ee
where the parameter $\delta^2$ determines the matrix
element 
\be
\langle \pi(p) |g_s\bar{d}\tilde{G}_{\alpha\mu}\gamma^\alpha u|0 \rangle=
i\delta^2f_\pi p_\mu ~.
\label{delta}
\ee
The nonasymptotic corrections to (\ref{phi4}) are not shown 
for brevity. At $Q^2=q^2$, the integrals over wave functions in 
(\ref{Ffinal}) convert into normalization factors
and the light-cone OPE is reduced to the short-distance expansion.    
The amplitude $F^{\gamma^{*} \pi}_{QCD}$ then simplifies:
\be
F^{\gamma^{*} \pi}_{QCD}(Q^2,Q^2)= \frac{\sqrt{2}f_\pi}{3Q^2}\left(1-
\frac89\frac{\delta^2}{Q^2} \right),
\label{QQlimit}
\ee
coinciding with the result of the short-distance expansion 
obtained in \cite{NSVZ}. 

The $O(\alpha_s)$ corrections to $F^{\gamma^{*} \pi}_{QCD}$ 
are beyound the scope of the present paper. 
Nevertheless, a few comments are in order. The perturbative 
$\alpha_s$--correction to the leading twist 2 term (\ref{leading}) 
has been calculated in \cite{rad}. 
One of the relevant diagrams is shown in Fig. 1c.
In our approach,  the account  of this effect 
requires a calculation of the imaginary part of the 
$O(\alpha_s)$-amplitude obtained in \cite{rad}. 
Simultaneously, the scale-dependence of the wave function (\ref{phipi}) 
should be taken into account in the next-to-leading order. 
The perturbative correction to the twist-4 contribution 
is unknown but is most likely inessential.
Furthermore, one has to take into account the $O(\alpha_s)$
contributions of four-quark operators  
to $F^{\gamma^{*} \pi}_{QCD}$. They were studied in \cite{BF} 
and in \cite{Gorsky}. However, the results differ, calling for a
new, independent calculation. The nonlocal four-quark matrix elements
have been approximated by factorizing two quark operators 
and taking their vacuum average $\langle \bar q q \rangle$. 
The remaining two operators then form a pion 
wave function of twist 3. One of the relevant diagrams is 
shown in Fig. 1d. Schematically, the corresponding 
correction to $F^{\gamma^{*}\pi}_{QCD}$ is 
\be
F^{\gamma^{*}\pi}_{(d)}(Q^2,q^2)
\sim \frac{\alpha_s \langle \bar q q \rangle}{Q^2q^2}\int\limits_0^1 
\frac{du~\varphi_{tw3}(u)}{Q^2(1-u) +q^2u}~.
\label{qqbar}
\ee
The divergence of (\ref{qqbar}) at $q^2\to 0$ clearly signals 
that a truncated light-cone OPE is not applicable at small $q^2$, even if 
$Q^2$ is large. In the full answer, this and similar divergences 
should cancel with  additional nonperturbative contributions 
corresponding to long-distance interactions of the photon. 
The latter can be taken into account by introducing the photon light-cone 
wave function.
For the short-distance OPE, such cancellation was studied in \cite{RR}.
The approach used here avoids this problem, because  
the hadronic dispersion relation is approximated by the light-cone  OPE  
at sufficiently large $q^2$, where the terms similar to (\ref{qqbar}) 
are suppressed.

It remains now to  
substitute in (\ref{lcsr}) and (\ref{answer}) 
the obtained expression (\ref{rhoqcd}) for 
the imaginary part $\mbox{Im}F^{\gamma^{*}\pi}_{QCD}$. 
Returning to the integration 
variable $u$  one finally obtains 
the $\gamma^* \rho \to \pi ^0$ form factor
\be
F^{\rho\pi}(Q^2) = \frac{f_\pi}{3f_\rho}V(Q^2,M^2)~,
\label{rhofinal}
\ee
and the $\gamma^* \gamma \to \pi^0$ form factor
\be
Q^2F^{\gamma\pi}(Q^2) 
= \frac{\sqrt{2}f_\pi}3\Bigg( \frac{Q^2}{m_\rho^2}
V(Q^2,M^2) + H(Q^2)\Bigg)~,
\label{pifinal}
\ee
where 
\be
V(Q^2,M^2) = \int^1_{\frac{Q^2}{s_0+Q^2}}\frac{du}u
\Bigg(\varphi_\pi (u)+\frac{u}{Q^2}\frac{d \varphi^{(4)}(u)}{du}\Bigg)
\exp\left(-\frac{Q^2(1-u)}{uM^2} +\frac{m_\rho^2}{M^2}\right)
\label{V}
\ee
and 
\be
H(Q^2) = \int_0^{\frac{Q^2}{s_0+Q^2}}\frac{du}{1-u}
\Bigg(\varphi_\pi(u)+\frac{u}{Q^2}\frac{d\varphi^{(4)}(u)}{du}\Bigg)~.
\label{H}
\ee

One should emphasize that the light-cone sum rule (\ref{rhofinal}) 
takes into account soft (end-point) contributions 
to the $\gamma^*\rho \to \pi$ form factor 
yielding $F^{\rho\pi}(Q^2) \sim 1/Q^4$ at $Q^2 \to \infty$ 
( for a more general discussion see \cite{BH}). 
In order to account for the hard-gluon exchange mechanism, which becomes
important at large momentum transfer, one should  
include the perturbative $\alpha_s$-correction in the sum rule. 

In the dispersion relation (\ref{pifinal}), the resonance part 
proportional to $V(Q^2)$ vanishes at $Q^2 \to \infty $ 
and $F^{\gamma\pi}(Q^2) \sim 1/Q^2$, in accordance with (\ref{leading}). 
At moderate $Q^2 \sim$ 1 GeV$^2$, the contributions from 
the vector meson and higher states  
are of the same order.

\section{Numerical results}
In order to proceed to the numerical analysis of the 
sum rule (\ref{rhofinal}) and relation (\ref{pifinal}), one has 
to specify the input.  
We take $f_\pi=132 $ MeV, $m_\rho= 770 $ MeV and $f_\rho= 216$ MeV.
The latter value is obtained  from (\ref{formrho})
and the $\rho^0 \to e^+ e^-$ width \cite{PDG}. 
The threshold parameter $s_0=1.5 $ GeV$^2$ is determined 
from the two-point sum rule in the $\rho$-meson channel \cite{SVZ}. 
The value of $\delta^2(1~\mbox{GeV})= 0.2$ GeV$^2$  
has been estimated from the corresponding sum rules in \cite{NSVZ,CZdelta}.
Furthermore, we consider three different approximations 
for the twist 2 wave function (\ref{phipi}): the 
asymptotic wave function ($a_n=0$), the CZ-wave function \cite{CZ} 
($a_2(\mu_0)=2/3$, $a_{n>2} =0$) and the BF-wave function 
\cite{BF} ($a_2(\mu_0)=2/3$, $a_4(\mu_0)=0.43$, 
$a_{n > 4}=0$), where $\mu_0 =0.5~\mbox{GeV}$. The nonasymptotic corrections to the twist 4 
wave functions entering (\ref{tw4wf})
have been roughly estimated in \cite{BF2}. Including them,
one obtains negligible changes of the numerical results. 
Hence, uncertainties of these corrections play no role here.  
Finally, the leading-order evolution
of $\varphi_\pi(u,\mu)$  and $\delta^2(\mu)$ is taken into account 
assuming $\mu=\sqrt{Q^2}$. 

In Fig. 2, the form factor $F^{\rho\pi}(Q^2)$
calculated from (\ref{rhofinal}) with the asymptotic 
$\varphi_\pi(u)$,  is plotted as a function of the Borel mass parameter $M$.
In light-cone sum rules,
the correlation function  is expanded in inverse powers of $uM^2$,
where $u$ is the light-cone momentum fraction, that is the
integration variable in (\ref{V}). To obtain suitable intervals of 
$M$ in (\ref{rhofinal}), we adopt $M^2= M^2_{2pt}/\langle u \rangle$,  
where $M_{2pt}$ is the Borel parameter of the two-point sum rule 
in the $\rho$-channel, and calculate the average value $\langle u \rangle $  
at each $Q^2$ separately. We then take 
$0.5< M^2_{2pt} < 0.8 ~\mbox{GeV}^2$, according to \cite{SVZ}.
The resulting interval of $M^2$ is shifting from 
$0.9 - 1.6 ~\mbox{GeV}^2$ at $ Q^2 \sim 
1 ~ \mbox{GeV}^2$ to $0.5 - 0.9 ~\mbox{GeV}^2$
at $Q^2=10~ \mbox{GeV}^2$. Within all these intervals, 
the twist 4 part of the light-cone sum rule 
does not exceed 35 \% and, simultaneously, the contribution from higher states 
estimated from duality is smaller than 40 \%.
At $Q^2 > 1$ GeV$^2$, the predicted form factor $F^{\rho\pi}(Q^2)$ 
is reasonably stable under variations of the Borel parameter in the 
adopted ranges. At $Q^2 < 1$ GeV$^2$, 
the sum rule (\ref{rhofinal}) becomes unstable signaling that 
one approaches too close to the physical region in the $\rho$ - channel.

Fig. 3 illustrates the sensitivity  of $F^{\rho \pi}(Q^2)$ 
(at $M^2_{2pt}=0.7$ GeV$^2$) 
to the choice of  nonasymptotic coefficients in $\varphi_\pi(u)$. 
We see that at $Q^2 \sim$ 10 GeV$^2$, the difference between the form 
factors calculated with the asymptotic 
wave function and with the CZ or BF wave functions 
is quite substantial. The observed sensitivity to the 
nonasymptotic effects is due to the fact that at large $Q^2$, 
the integration over $u$ in the light-cone sum rule 
is restricted to the end-point region, 
approximately, to the interval $1-s_0/Q^2 < u < 1$. 
In this region, the integrals over nonasymptotic parts 
of the wave function (\ref{phipi}) proportional to the Gegenbauer 
polynomials are of the same order as the integrated asymptotic part.
The twist 4 contribution to (\ref{rhofinal}) is between 
35\% and 10\% at $1~\mbox{GeV}^2 < Q^2 < 10~\mbox{GeV}^2$. As already 
mentioned, this contribution is dominated by asymptotic wave functions 
and therefore has a small uncertainty. We conclude that measurements of the 
$\gamma^* \rho \pi $ and $\gamma^* \omega \pi $ 
form factors at momentum transfer of order of a few GeV$^2$ 
can  indeed be used to discriminate between 
various approximations for the twist 2 wave function $\varphi_\pi(u)$.

The form factor $F^{\gamma\pi}(Q^2)$ is calculated from 
the relation (\ref{pifinal}) with the same numerical input. 
In Fig. 4, it is plotted taking the asymptotic $\varphi_\pi(u)$
and $M^2_{2pt}=0.7$ GeV$^2$. 
The twist 2 and 4 contributions are shown separately. 
We see a nontrivial
$Q^2$-dependence of this form factor. 
At  $Q^2 < 10$ GeV$^2$, it noticeably deviates 
from the asymptotic limit $ Q^2 F^{\gamma\pi}(Q^2)\to \sqrt{2}f_\pi$ .
Fig. 5 shows the predictions on  $F^{\gamma\pi}(Q^2)$ 
obtained with other choices of the twist 2 wave function. 
Starting from $Q^2 \simeq 3 - 4$ GeV$^2$,
the role of the nonasymptotic part is quite essential. 

The main uncertainty of the obtained predictions  
is due to the neglect of the perturbative $\alpha_s$-correction and   
will be removed, once this correction is taken into account.
The role of four-quark contributions such as (\ref{qqbar}), 
which are suppressed by extra powers of photon virtualities 
and $\alpha_s$, cannot be important at $Q^2 > $ 1 GeV$^2$. 
In order to estimate the accuracy of the leading-order 
approximation in $\alpha_s$ adopted here, the Borel parameter $M_{2pt}^2$ 
was varied within $0.5- 0.8 $ GeV$^2$  and the threshold parameter 
$s_0$ within $1.3 - 1.8 $ GeV$^2$. The resulting variations 
of $F^{\rho\pi}(Q^2)$ around the predictions shown in Fig. 3 
are $\pm 5 \%$  and $\pm 10\%$, respectively, almost independent 
of $Q^2$. The corresponding variations of $F^{\gamma\pi}(Q^2)$ are
$\pm 3\%$ and $\pm 2 \%$ at $Q^2 \sim 1 $ GeV$^2$, 
and become negligibly small at larger $Q^2$.   
An additional uncertainty is connected with 
the choice of the normalization scale $\mu$ which is 
somewhat arbitrary in the absence of $\alpha_s$-correcitons.  
Taking a $Q^2$-independent scale $\mu = 1$ GeV, which is of order
of the Borel parameter, does not change the results at 
$Q^2 \sim 1$ GeV$^2$, but yields a 25\% (10\%) increase 
of $F^{\rho\pi}$ ($F^{\gamma\pi}$) at $Q^2 \sim $ 10 GeV$^2$
in the case of the CZ and BF wave functions. 
The inclusion of the perturbative $\alpha_s$-correction will 
certainly weaken this scale-dependence. 

To have a more complete account  of uncertainties of the method, 
one also has to assess the accuracy 
of the dispersion relation (\ref{disp}). In this relation, the isospin
symmetry is assumed neglecting  $\rho-\omega$ mixing and 
adopting (\ref{formf}) and (\ref{formrho}). This is consistent with the 
isospin-symmetry limit of the light-cone OPE of the amplitude 
$F^{\gamma^* \pi}_{QCD}(Q^2,q^2)$. In addition, we adopt 
the zero-width approximations for $\rho$ and $\omega$.
To clarify the sensitivity of form factors $F^{\rho \pi}$ 
and $F^{\gamma \pi}$ to these approximations, 
the resonance term in (\ref{disp}) has been modified to a 
finite-width Breit-Wigner form: 
\be
\label{BW}
\frac{\sqrt{2}
f_\rho F^{\rho \pi}(Q^2)}{m_{\rho}^2+q^2} \to 
\frac1{\sqrt{2}\pi}\sum_{V=\rho,\omega}~\int\limits_{4m_\pi^2}^{s_0}
ds ~\frac{m_V\Gamma_Vf_V F^{V\pi}(Q^2)}{[(m_V^2-s)^2+m_V^2\Gamma_V^2]
(s+q^2)}~.
\ee
substituting the experimental values \cite{PDG} of $\Gamma_\rho= 151$ MeV, 
$\Gamma_\omega= 8$ MeV, $m_\omega=782$ MeV, and $f_\omega \simeq 1/3(0.9)f_\rho$, and retaining $F^{\omega \pi}(Q^2)\simeq  3 F^{\rho \pi}(Q^2)$. 
Numerically, the substitution (\ref{BW}) yields a 12\% (6\%) 
increase of $F^{\rho\pi}$ ($F^{\gamma\pi}$).  
We use the magnitude of this change as a rough estimate of the combined 
uncertainty due to the resonant part in the dispersion relation
(\ref{disp}). 

Finally, for convenience, the obtained results on $\gamma^*\rho\to \pi$ 
and $\gamma^*\gamma\to \pi^0$ form factors in the region 
$1 < Q^2 < 10$ GeV $^2$ have been fitted
to the parametrizations 
\be
F^{\rho\pi}(Q^2)Q^4=\frac{A^{\rho\pi}}{ 1+\frac{B^{\rho\pi}}{Q^2}+\frac{C^{\rho\pi}}{Q^4}} ~,
\ee
and
\be
F^{\gamma\pi}(Q^2)Q^2=\frac{A^{\gamma\pi}}{1+\frac{B^{\gamma\pi}}{Q^2}}
\ee
with
\be
A^{\rho\pi}=0.92\pm 0.2 ~(1.94 \pm 0.55),~~  
B^{\rho\pi}=3.96~(2.27)\mbox{GeV}^2,~~C^{\rho\pi}=2.48~(13.5)\mbox{GeV}^4  
\ee
and 
\be
A^{\gamma\pi}=0.186\pm 0.02~(0.242 \pm 0.04),~~  
B^{\rho\pi}=0.875~(1.385)\mbox{GeV}^2.
\ee
The numerical values of the above parameters 
correspond to the asymptotic (CZ) choice of the pion light-cone wave function.
The quoted normalization errors (conservatively) take into 
account  the estimated theoretical uncertainties of the 
leading-order approximation in $\alpha_s$ considered in this analysis.

\section{Conclusion}
In this paper, the $\gamma^*\rho \to \pi$ and 
$\gamma^*\gamma \to \pi^0$ form factors 
have been calculated using the light-cone OPE, the dispersion relation 
and the quark-hadron duality in the $\rho$-meson channel.  
The main results are in
(\ref{rhofinal}) - (\ref{H}),
expressing $F^{\rho\pi}(Q^2)$  and $F^{\gamma\pi}(Q^2)$, respectively, 
in terms of light-cone wave functions of the pion. 
At $Q^2$ of order of a few  GeV$^2$, the numerical predictions 
on both form factors are sensitive to nonasymptotic 
effects in the twist 2 wave function $\varphi_\pi(u)$.

In Fig. 5, the obtained results for 
the form factor $F^{\gamma\pi}(Q^2)$ 
are compared with the new CLEO data \cite{CLEO} and with the earlier 
CELLO data \cite{CELLO}. This comparison supports 
the asymptotic form of the wave function $\varphi_\pi(u)$. 
More definite quantitative conclusions can be 
made after including perturbative corrections in our analysis.
Note that the $\gamma^*\rho\to \pi$  form factor can also be measured, 
e.g. by extracting the one-pion exchange in the electroproduction 
of $\rho,\omega$ mesons \cite{rhopigamma}. 

In Fig. 3, our prediction on $F^{\rho\pi}(Q^2)$ 
is compared with the results of other 
calculations. In  \cite{BH}, a light-cone sum rule 
for the $\gamma^*\rho_\perp\to \pi$  transition  form factor 
has been obtained from a correlation 
function of two currents, $j_\mu$ and 
$\bar d \sigma_{\mu\nu}u $ 
($\rho_\perp$ is a $\rho$-meson with the helicity $\lambda=\pm1$). 
The leading contribution 
to this sum rule is generated by the twist 3 wave function
of the pion. The higher-twist terms are not known, hence, the achieved 
accuracy is not high. Therefore, only a crude agreement with our prediction 
obtained with the asymptotic $\varphi_\pi(u)$ can be expected. 
Fig. 3 also shows the $\gamma^*\rho\to \pi$ form factor obtained 
\cite{EK} from the three-point sum rule 
\footnote{ we take into account that in 
\cite{EK} the normalization of the form factor contains a factor 
$\sqrt{4\pi\alpha}$.} in the region 
$ Q^2 = 0.5-3 $ GeV$^2$ . The three-point sum rule prediction is in a 
good agreement with our result obtained with the CZ and BF wave 
functions. The latter result also agrees with 
the form factor $F^{\rho\pi}(Q^2)$  calculated  in the  
relativistic quark model \cite{Aznauri} 
in the same region. Furthermore, 
the form factor $F^{\rho\pi}(Q^2)$ obtained  
in the light-front constituent quark model \cite{Narod} 
at  $Q^2 = 1-8 $ GeV$^2$, is quite close 
to our prediction obtained with the asymptotic $\varphi_\pi(u)$.

Turning to the $\gamma^*\gamma \to \pi^0$ transition, we see from
Fig. 5 that at $Q^2 > 1 $ GeV$^2$ the relation (\ref{pifinal}) 
is in a good numerical agreement with the simple interpolation formula 
\be
F^{\gamma\pi}(Q^2)= \frac{\sqrt{2}f_\pi}{4\pi^2f_\pi^2+ Q^2}~,
\label{BLint}
\ee
suggested in \cite{BL}, if  the asymptotic $\varphi_\pi(u)$ is adopted. 
In \cite{RR}, the form factor $F^{\gamma\pi}(Q^2)$ 
was calculated using 3-point correlation function, short-distance OPE 
and  QCD sum rule in the pion channel. The long-distance interaction 
of the small virtuality photon was taken into 
account introducing bilocal correlators, 
employing duality and light-cone wave functions.
After that, $F^{\gamma\pi}(Q^2)$  has been obtained in terms 
of a combined nonperturbative input 
including quark/gluon condensates and light-cone wave functions
of $\rho$ -meson and photon. Numerically, the result of 
\cite{RR} is close to the interpolation formula (\ref{BLint}) 
and therefore also to our prediction for the 
$\gamma^*\gamma \to \pi^0$ 
form factor obtained with the asymptotic
wave function of the pion.

\bigskip

{\Large \bf Acknowledgements}.

\bigskip

I am grateful to I. Aznaurian, V. Braun, S. Brodsky, V. Eletsky, 
I. Halperin, R. R\"uckl, M. Shifman, A. Vainshtein  and O. Yakovlev 
for useful discussions and comments concerning various aspects of this study. 
My interest to the two-photon transition into pion was initiated
by communication of V. Savinov. This work 
is supported by the German Federal Ministry for 
Research and Technology (BMBF) under contract number 05 7WZ91P (0).
The main part of the work was completed during the Workshop 
on Light-Cone QCD in Lutsen, Minnesota. I am grateful to J. Hiller for
hospitality and to TPI, University of Minnesota for travel support.

\newpage


\begin{figure}[htb]
\centerline{
\epsfig{bbllx=24pt,bblly=65pt,bburx=525pt,%
bbury=620pt,file=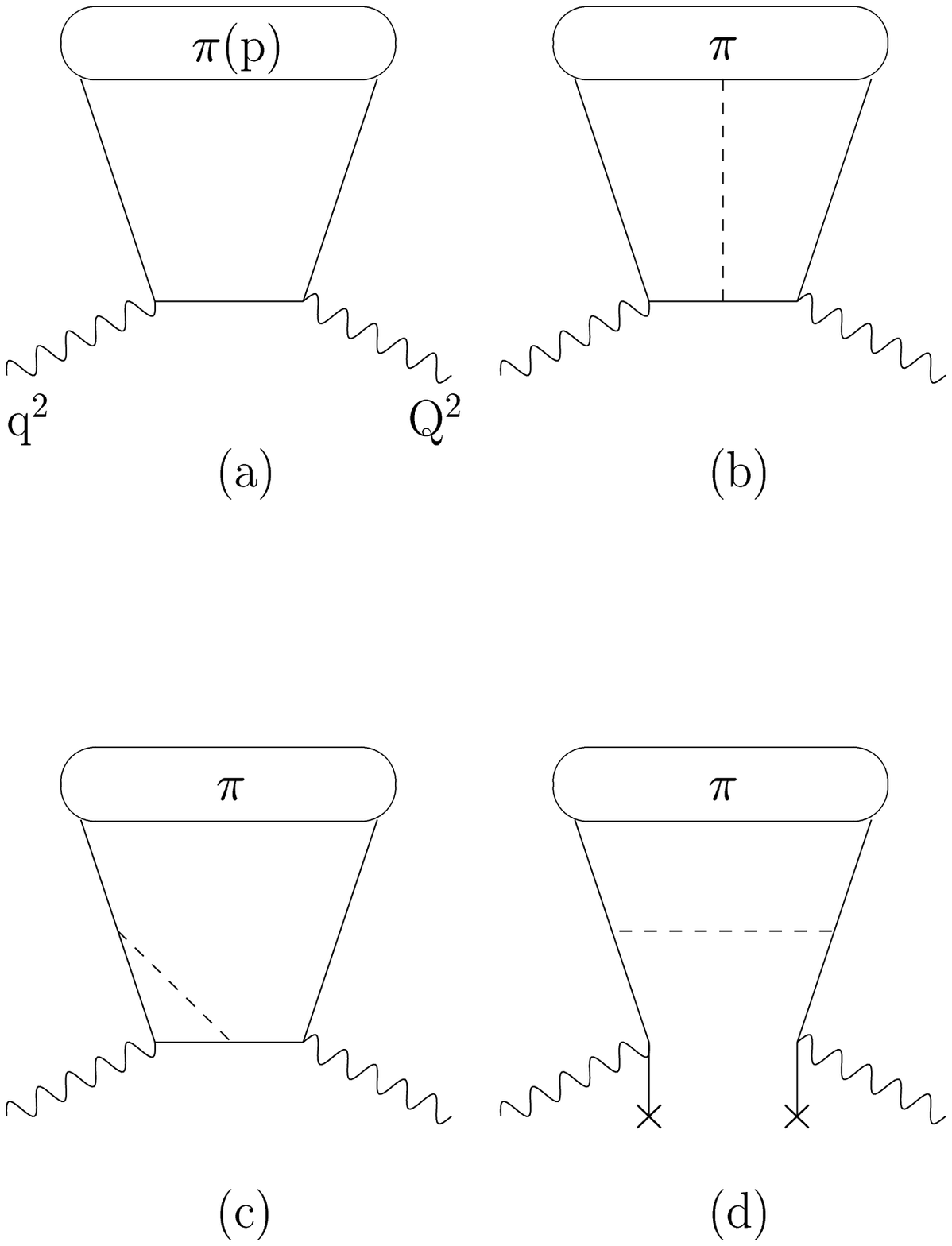,scale=0.9,%
,clip=}
}
\caption{\it 
Diagrams corresponding to the 
light-cone OPE of the amplitude (\ref{ampl}).
Solid lines represent quarks, dashed lines gluons, wavy lines 
electromagnetic currents. The ovals denote light-cone wave
functions of the pion.}
\end{figure}

\begin{figure}[htb]
\centerline{
\epsfig{bbllx=45pt,bblly=240pt,bburx=426pt,%
bbury=656pt,file=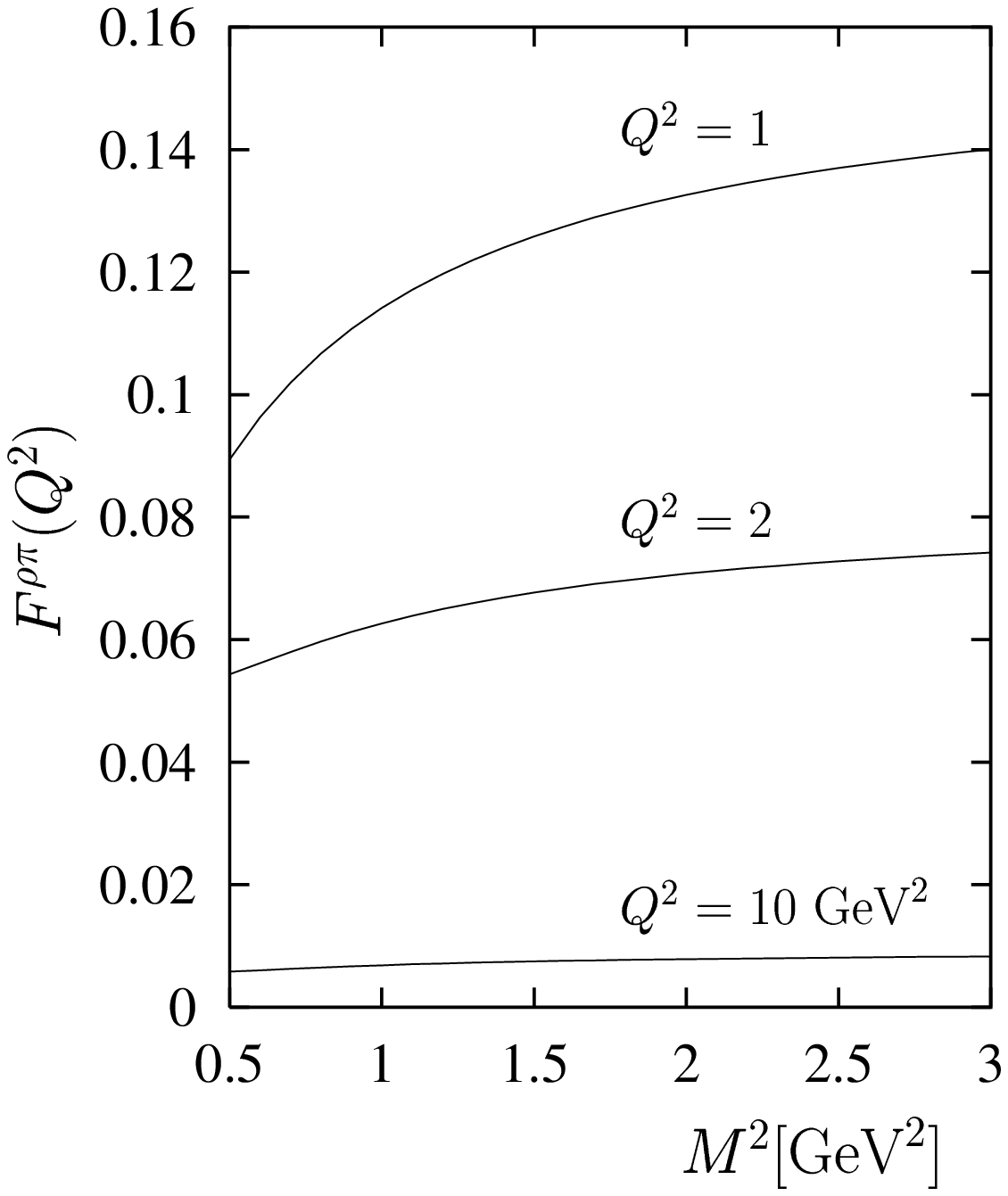,scale=0.9,%
,clip=}
}
\caption{\it 
Form factor of the $\gamma^*\rho \to \pi$ transition obtained 
from the light-cone sum rule as a function of the Borel 
parameter at different values of the momentum transfer.
} 
\end{figure}

\begin{figure}[htb]
\centerline{
\epsfig{bbllx=79pt,bblly=256pt,bburx=575pt,%
bbury=550pt,file=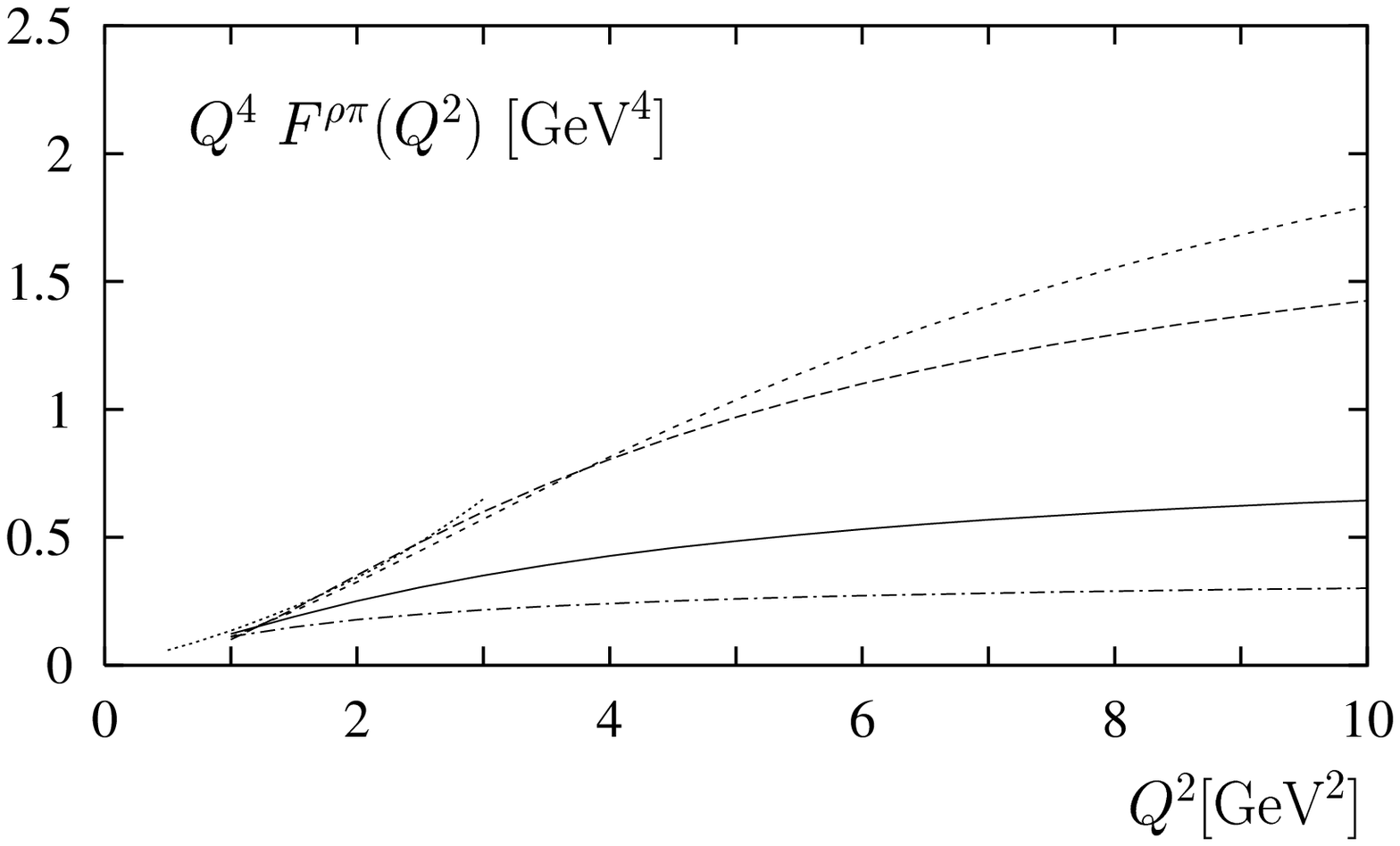,scale=0.9,%
,clip=}
}
\caption{\it 
$\gamma^*\rho \to \pi$  form factor calculated 
from the light-cone sum rule with the asymptotic 
pion wave function (solid), with the 
CZ wave function (long-dashed) and with the BF wave 
function (short-dashed), in comparison with the
predictions of the  
three-point QCD sum rule (dotted) \cite{EK}, and 
light-cone sum rule for the $\gamma^*\rho_\perp\to \pi$
form factor \cite{BH} (dash-dotted).} 
\end{figure}

\begin{figure}[htb]
\centerline{
\epsfig{bbllx=63pt,bblly=250pt,bburx=576pt,%
bbury=546pt,file=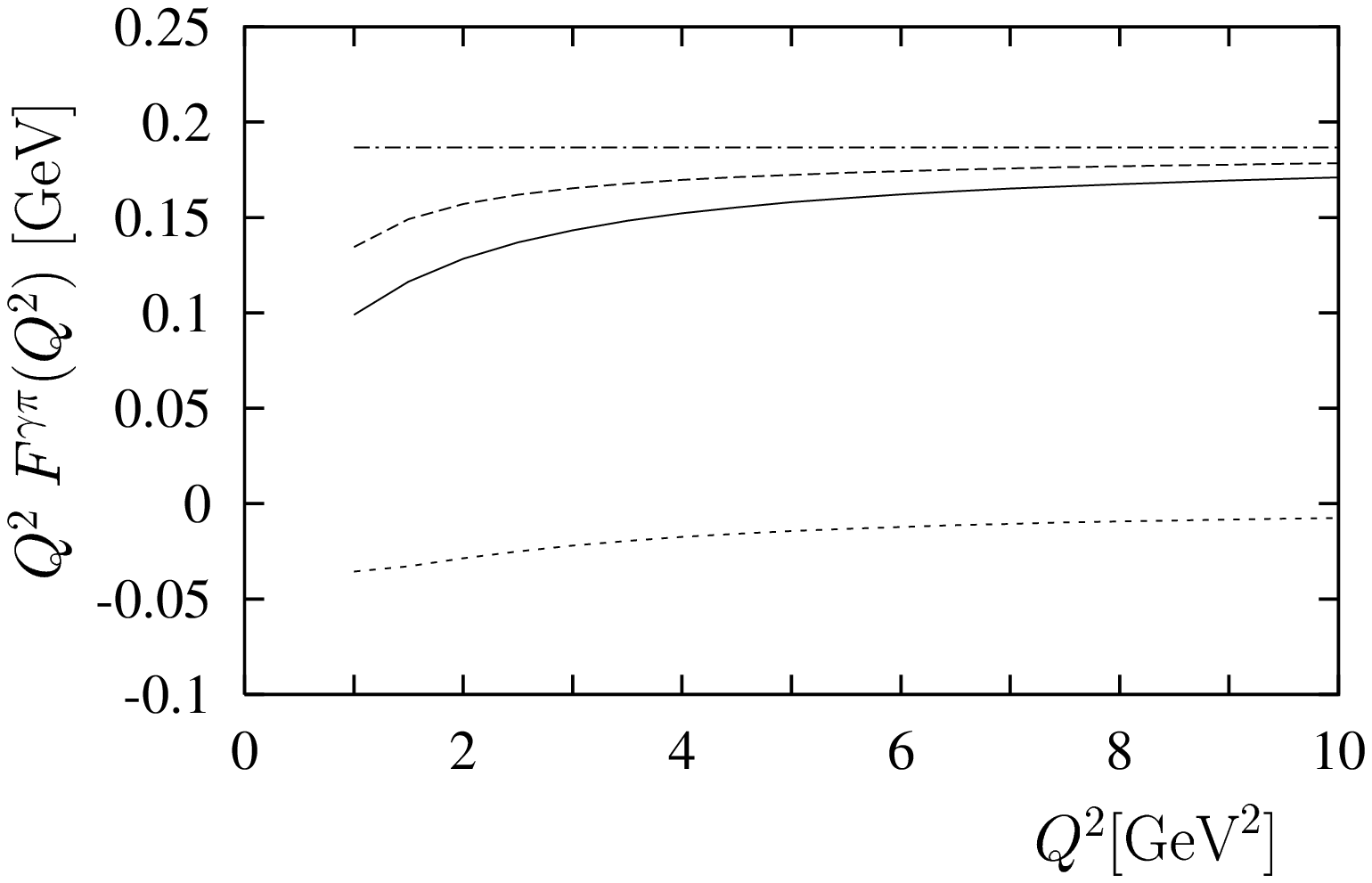,scale=0.9,%
,clip=}
}
\caption{\it 
Form factor of the $\gamma^*\gamma \to \pi^0$ transition 
calculated 
from the relation (\ref{pifinal}) with the asymptotic
wave function of the pion (solid), twist 2 (dashed), twist 4 (dotted) 
contributions and the $Q^2\to \infty$ limit (dash-dotted).} 
\end{figure}

\begin{figure}[htb]
\centerline{
\epsfig{bbllx=38pt,bblly=245pt,bburx=510pt,%
bbury=546pt,file=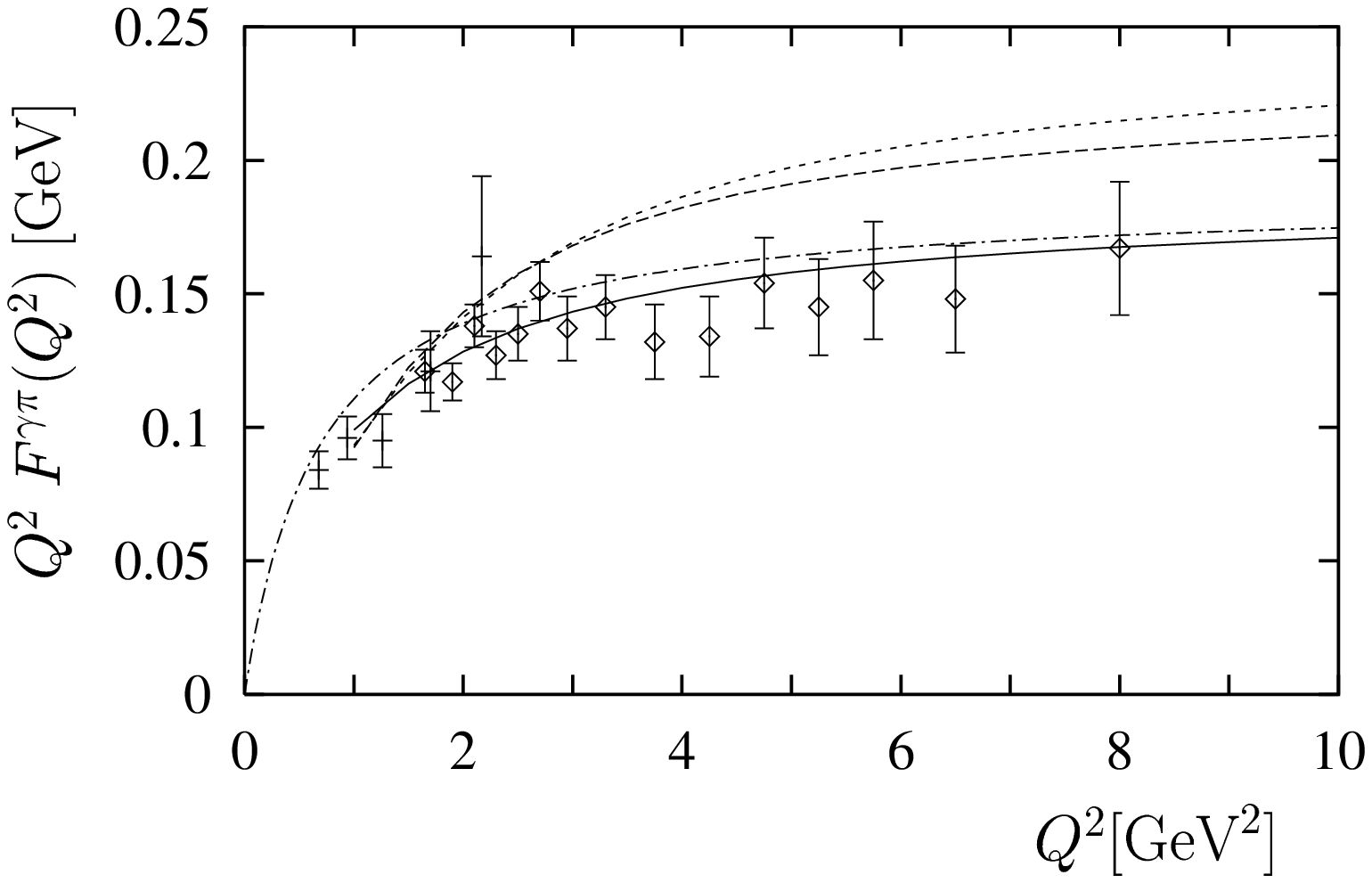,scale=0.9,%
,clip=}
}
\caption{\it 
Form factor of the $\gamma^*\gamma \to \pi^0$ transition 
calculated with the asymptotic 
(solid), CZ (long dashed) and BF ( short-dashed) wave 
function of the pion in comparison with the experimental 
data points \cite{CLEO,CELLO} and with the  
interpolation formula (\ref{BLint}) from \cite{BL} (dash-dotted).} 
\end{figure}


\newpage
\begin{thebibliography}{99}

\bibitem{BL}
 G.P. Lepage and S.J. Brodsky,
  Phys. Lett. B87 (1979) 359;
  Phys. Rev. D22 (1980) 2157; Phys. Rev. D24 (1981) 1808.

\bibitem{ER}
V.L. Chernyak and A.R. Zhitnitsky,
JETP Lett.  25 (1977) 510; 
Sov. J. Nucl. Phys. 31 (1980) 544;   


A.V. Efremov and A.V. Radyushkin,
  Phys. Lett.  B94 (1980) 245;
  Teor. Mat. Fiz.  42 (1980) 147.

\bibitem{BBK} 
I.I. Balitsky, V.M.~Braun and A.V.~Kolesnichenko,
 Sov. J. Nucl. Phys.  44  (1986) 1028;
 Nucl. Phys. B312 (1989) 509.
\bibitem{BF}
V.M. Braun and I.E. Filyanov, Z. Phys. C44 (1989) 157.

\bibitem{CZ90}
V.L. Chernyak and I.R.Zhitnitsky, 
 Nucl. Phys. B345 (1990) 137.
 
\bibitem{BKR} V.M. Belyaev, A. Khodjamirian and  R. R\"uckl, 
Z. Phys. C 60 (1993) 349.  

\bibitem{BBKR} V.M. Belyaev, V.M.~Braun, 
A. Khodjamirian and  R. R\"uckl, 
Phys. Rev D 51 (1995) 6177. 

\bibitem{BH}
V. Braun, I. Halperin, Phys. Lett. B328 (1994) 457. 

\bibitem{CLEO} 
V. Savinov, Contribution to the Intern. Conference Photon 97, 
Egmond aan Zee, Netherlands, (1997), hep-ex/9707028;\\ 
CLEO Collaboration (J. Gronberg et al.),
preprint CLNS-97-1477 (1997), hep-ex/9707031; 


\bibitem{RR} 
A.V. Radyushkin, R.T. Ruskov,  Nucl. Phys. B481 (1996) 625. 


\bibitem{RM} 
I.V. Musatov, A.V. Radyushkin,  Phys. Rev. D56 (1997) 2713.


\bibitem{ShifSmil} A.V. Smilga, M.A. Shifman, 
Sov. J. Nucl. Phys. 37 (1983) 958.
 
\bibitem{GIKO}
A.S. Gorski, B.L. Ioffe, A.Yu. Khodjamirian and A.G. Oganessian,
Z. Phys. C44 (1989)  523 .     

\bibitem{Gorsky}
A.S.~Gorsky, Sov. J. Nucl. Phys. 41 (1985) 1008;
{\it ibid.} 45 (1987) 512; {\it ibid.} 50 (1989) 498 .

\bibitem{BF2}V.M. Braun and I.E. Filyanov, Z. Phys. C48 (1990) 239.


\bibitem{NSVZ}  
V.A. Novikov, M.A. Shifman, A.I. Vainshtein and V.I. Zakharov, Nucl.
Phys. B237 (1984) 525.

\bibitem{rad} F. del Aguila and M.K. Chase,  
Nucl. Phys. B193  (1981) 517;\\
E. Braaten, Phys. Rev. D28 (1983) 524;\\
E.P. Kadantseva, S.V. Mikhailov and A.V. Radyushkin,
 Sov. J. Nucl. Phys. 44  (1986) 326.


\bibitem{PDG}  Particle Data Group, Phys. Rev. D54 (1996) 1.



\bibitem{SVZ}  
M.A. Shifman, A.I. Vainshtein and V.I. Zakharov, Nucl.
Phys. B147 (1979) 385, 448.

\bibitem{CZdelta} 
V.L. Chernyak, A.R. Zhitnitsky and I.R. Zhitnitsky, Sov. J. 
Nucl. Phys. 38 (1983) 645.


\bibitem{CZ}
V.L. Chernyak and A.R. Zhitnitsky,
Phys. Rep.  112 (1984) 173. 


\bibitem{CELLO} CELLO Collaboration ( H.-J. Behrend et. al),
Z. Phys. C49 (1991) 401.  

\bibitem{rhopigamma}
I.G. Aznauryan, Phys. Atom. Nucl. 60 (1997) 584.

\bibitem{EK} 
V.L. Eletskii, Ya.I. Kogan, Z. Phys. C20 (1983) 357.
 
\bibitem{Aznauri} I.G. Aznauryan and K. A. Oganessyan, 
Phys. Lett. B249 (1990) 309.

\bibitem{Narod}
F. Cardarelli, I.L. Grach, I. Narodetskii, G. Salme and S. Simula,
Phys. Lett. B359 (1995) 1. 



\end{thebibliography}
\end{document}